\documentclass[preprint,electronic]{vgtc}             

  \pdfoutput=1\relax                   
  \usepackage{hyperref}
  \usepackage{graphicx}                
  \DeclareGraphicsExtensions{.pdf,.png,.jpg,.jpeg} 

\graphicspath{{figures/}{pictures/}{images/}{./}} 

\usepackage{microtype}                 
\PassOptionsToPackage{warn}{textcomp}  
\usepackage{textcomp}                  
\usepackage{mathptmx}                  
\usepackage{times}                     
\usepackage{cite}                      
\usepackage{booktabs}                  

\usepackage[svgnames]{xcolor}
\usepackage{paralist}

\usepackage{listings} 

\definecolor{darkblue}{rgb}{0.0,0.0,0.6}
\colorlet{orangeb}{orange!80!black}
\def\noprint#1{}

\colorlet{punct}{red!60!black}
\definecolor{background}{HTML}{EEEEEE}
\definecolor{delim}{RGB}{20,105,176}
\colorlet{numb}{magenta!60!black}

\lstdefinelanguage{json}{
    basicstyle=\normalfont\ttfamily,
    numbers=left,
    numberstyle=\scriptsize,
    stepnumber=1,
    numbersep=8pt,
    showstringspaces=false,
    breaklines=true,
    literate=
     *{0}{{{\color{numb}0}}}{1}
      {1}{{{\color{numb}1}}}{1}
      {2}{{{\color{numb}2}}}{1}
      {3}{{{\color{numb}3}}}{1}
      {4}{{{\color{numb}4}}}{1}
      {5}{{{\color{numb}5}}}{1}
      {6}{{{\color{numb}6}}}{1}
      {7}{{{\color{numb}7}}}{1}
      {8}{{{\color{numb}8}}}{1}
      {9}{{{\color{numb}9}}}{1}
      {:}{{{\color{punct}{:}}}}{1}
      {,}{{{\color{punct}{,}}}}{1}
      {\{}{{{\color{delim}{\{}}}}{1}
      {\}}{{{\color{delim}{\}}}}}{1}
      {[}{{{\color{delim}{[}}}}{1}
      {]}{{{\color{delim}{]}}}}{1},
}

\providecommand{\highlight}[1]{`#1'} 

\onlineid{9717}

\vgtccategory{VAHC}

\ieeedoi{10.1109/VAHC47919.2019.8945032}
\preprinttext{\parbox{0.9\textwidth}{$\copyright$ 2019~IEEE. This is the author's version of the article that has been published in the proceedings of IEEE VAHC. The final version of this record is available at: \href{https://doi.org/10.1109/VAHC47919.2019.8945032}{\color{blue}10.1109/VAHC47919.2019.8945032}}}

  \hypersetup{
   pdfpagemode=UseNone,
   pdftitle={},
   pdfauthor={},
   pdfsubject={},
   pdfkeywords={},
   pageanchor=true,
   plainpages=false,       
   hypertexnames=false,    
   bookmarksnumbered=false,
   bookmarksopen=true,     
   bookmarksopenlevel=3,   
   pdfpagemode=UseNone,    
   pdfstartview=Fit,       
   pdfborder={0 0 0},      
   breaklinks=true,        
   colorlinks=false,       
   nesting=true}

  \renewcommand{\pdfbookmark}[3][]{}

\title{Towards a Structural Framework\\for Explicit Domain Knowledge in Visual Analytics}

\author{Alexander Rind\thanks{e-mail: \{firstname\}.\{lastname\}@fhstp.ac.at} %
\and Markus Wagner\footnotemark[1] 
\and Wolfgang Aigner\footnotemark[1]} 
\affiliation{\scriptsize
Center for Digital Health Innovation, Institute of Creative$\backslash{}$Media/Technologies\\
St.~P\"olten University of Applied Sciences, Austria}

\abstract{%
Clinicians and other analysts working with healthcare data are in need for better support to cope with large and complex data.
While an increasing number of visual analytics environments integrates explicit domain knowledge
as a means to deliver a precise representation of the available data,
theoretical work so far has focused on the role of knowledge in the visual analytics process.
There has been little discussion about how such explicit domain knowledge can be structured in a generalized framework.
This paper collects desiderata for such a structural framework
, proposes how to address these desiderata based on the model of linked data,
and demonstrates the applicability in a visual analytics environment for physiotherapy.%
} 

\CCScatlist{
  \CCScatTwelve{Human-centered computing}{Visu\-al\-iza\-tion}{Visu\-al\-iza\-tion theory, concepts and paradigms}{};
  \CCScatTwelve{Applied computing}{Life and medical sciences}{Health care information systems}{}
}

\begin{document}

\firstsection{Introduction}

\maketitle

To keep pace with the tremendously expanding volumes of complex and heterogeneous data,
experts in many domains such as healthcare need to apply high-performance data analysis methods.
Even though the sheer quantity demands automated analysis methods, this process cannot be automated completely,
since domain experts need to be in the loop to identify, correct, and disambiguate intermediate results \cite{wegner_1997_why}.
\emph{Visual analytics} (VA) intertwines interactive visual interfaces with automated data analysis methods in order to support humans in data analysis \cite{keim_2010_mastering,thomas_2005_illuminating}.
This allows for effectively distributing the workload of cognitive reasoning between human and machine  \cite{kirsh_2010_thinking,liu_2008_distributed}.
However, this endeavor is not straightforward as initial results from automated analysis are often trivial or irrelevant to the work of the domain experts \cite{mccurdy_2018_framework}.
Domain knowledge is needed to sift the relevant from the trivial.

Let us illustrate this with a hypothetical example:
The general practitioner Jane is treating Mary who suffers from diabetes for more than a decade.
Today, Mary's creatinine levels are elevated and Jane needs to adjust treatment in order to avert kidney damage.
She needs to consult Mary's electronic health record to check for past medication with possible side effects on the kidney.
However, the record is quite voluminous due to the long duration of treatment, comorbidities, lifestyle changes, pregnancy, and unrelated events such as seasonal colds.
Jane's medical experience helps her identify the relevant events,
but she wonders if this knowledge can be part of her VA environment
so that it automatically links medical findings to possible causes and available treatments.

Better integrating analysts' knowledge has been emphasized by the VA community as a central research challenge in the field \cite{andrienko_2018_model,chen_2005_top,chen_2009_data,keller_2005_visualizing,pike_2009_science}.
Consequently, an increasing number of VA environments integrate explicit knowledge,
which is knowledge that ``can be processed by a computer, transmitted electronically, or stored in a database'' \cite[p.\,617]{wang_defining_2009}.
Federico, Wagner et al. \cite{federico-wagner_2017_model} recently
coined the term \emph{knowledge-assisted visual analytics}
for such environments that have
``features to generate, transform, and utilize explicit knowledge'' \cite[p.\,92]{federico-wagner_2017_model}. 

The emerging integration of explicit knowledge in VA environments,
in particular through design studies that are grounded on the concrete needs of a target audience \cite{sedlmair_2012_design},
provides the opportunity to reflect current practice and develop general guidelines and theoretical models for knowledge in VA \cite{thomas_2005_illuminating}.
However, theoretical work in knowledge-assisted VA to date tended to focus on the role of knowledge in the VA process \cite{federico-wagner_2017_model,ribarsky_human-computer_2016,wang_defining_2009} rather than on the explicit knowledge itself.
Our field still lacks a generalized framework of how explicit domain knowledge can be structured, stored and made accessible to a VA environment.
The prospective value of generalizing the particularly designed mechanisms for explicit knowledge are threefold:
\begin{inparaenum}[\itshape (i)]
\item
it allows us to better understand and compare VA environments by their integration of explicit knowledge,
\item
it guides the development of future VA design studies, and
\item
it is a precondition for
the exchange of explicit knowledge between VA environments. 
\end{inparaenum}
This third point, in particular, envisions an analytics ecosystem in which explicit knowledge is a first-class artifact and not limited to the scope of a single tool but can be reused in all the tools needed to perform an activity, without bothering the human to transcribe it into each environment separately.

Therefore, 
this paper collects desiderata for a structural framework of explicit domain knowledge in VA (Sect.~\ref{sec:desiderata}).
These desiderata originate from reflective discussions of three design studies
that resulted in knowledge-assisted VA environments for different application domains
(internal medicine \cite{rind_2011_visualexploration}, gait rehabilitation \cite{wagner_2018_kavagait},
and IT security \cite{wagner_2017_kamas})
and the review of the scientific literature, which is summarized in Sect.~\ref{sec:relwork}.
Sect.~\ref{sec:model} proposes how to address these desiderata based on the model of linked data.
To demonstrate the applicability, Sect.~\ref{sec:case} presents
how explicit domain knowledge is integrated in the VA environment
KAVAGait \cite{wagner_2018_kavagait}
that supports physiotherapists in gait rehabilitation.

\section{Background and Related Work}
\label{sec:relwork}

Since ``Illuminating the Path'' \cite[p.\,35]{thomas_2005_illuminating}, 
incorporating prior domain knowledge and ``build[ing] knowledge structures'' has been on VA's agenda.
This is underscored by the pivotal position of knowledge in the VA process model by Keim et al.~\cite{keim_visual_2008,keim_2010_mastering}
and further process models such as the knowledge generation model by Sacha et al.~\cite{sacha_knowledge_2014}
and the visualization model by van Wijk \cite{wijk_value_2005}.
However, these process models do not differentiate
between knowledge in the human space and in the machine space.
Based on Wang et al.~\cite{wang_defining_2009}, Federico, Wagner et al.~\cite{federico-wagner_2017_model} delineate tacit knowledge that is exclusively available to human reasoning,
from explicit knowledge that can be leveraged by the VA environment.
How explicit knowledge is integrated into the VA process is formalized in several recent models by Wang et al.~\cite{wang_defining_2009}, Ribarsky et al.~\cite{ribarsky_human-computer_2016},
and Federico, Wagner et al.~\cite{federico-wagner_2017_model}.

Beyond the role of knowledge in the VA process,
only few works discuss the content and structure of explicit knowledge on a general level.
Andrienko et al. \cite{andrienko_2018_model} conceptualize domain knowledge as a model of a part of reality and provide definitions for different types of models
but they do not specify the form and medium how the model is represented.
Schulz et al.~\cite{schulz_2017_systematic} formalize data descriptors that include domain knowledge about data.
Tominski~\cite{tominski_2011_event-based} captures domain knowledge as event types that are specified using predicate logic.
Lammarsch at al.~\cite{lammarsch_2011_towards} propose a data structure for knowledge about temporal patterns leveraging the structure of time.
The generation of adapted visualizations which are based on ontological datasets and the specification of ontological mappings are treated by Falconer et al.~\cite{falconer_creating_2009}.
Therefore, they use the \highlight{COGZ} tool, converting ontological mappings in software transformation rules so that it describes a model which fits the visualization.
A similar approach for adapted visualizations is also followed by Gilson et al.~\cite{gilson_web_2008}, describing a general system pipeline which combines ontology mapping and probabilistic reasoning techniques. Thereby, they describe the automated generation of visualizations of domain-specific data from the web.
However, none of these approaches aim for a general framework.

The application of visualization techniques to \textbf{healthcare} has sparked a lot of interest to integrate knowledge.
Already the early LifeLines~\cite{plaisant_1998_lifelines} approach envisioned how domain knowledge is used to highlight relationships between events, which was then realized by a simple full-text search.
A number of approaches such as Midgaard \cite{bade_2004_connecting, aigner_2012_comparative} and QualizonGraph \cite{federico_2014_qualizon} enrich the display of time series of medical parameters with qualitative levels.
Thus, a period of critical conditions can be detected and visually highlighted, for example by color.
The ViTA-Lab environment \cite{klimov_2014_exploration} and its preceding work \cite{klimov_2010_intelligentvisualization, shahar_1999_intelligent} leverage complex temporal data abstractions for pattern discovery and provide a case study of longitudinal analysis of 22,000 diabetes patient records.
The Five W's \cite{zhang_2013_five} environment arranges events of the health record hierarchically based on the knowledge about diseases formalized in the ICD9 taxonomy.
Gnaeus \cite{federico_2015_gnaeus} is a guideline-based knowledge-assisted visualization of electronic health records for cohorts. 
Evidence-based clinical practice guidelines are sets of statements and recommendations used to improve health care by providing a trustworthy comparison of treatment options in terms of risks and benefits according to patient's status. 
The KAVAGait \cite{wagner_2018_kavagait} tool is a knowledge-assisted VA environment for clinical gait analysis that supports analysts during diagnosis and clinical decision making. Users can load, visualize and compare patient gait data containing \textit{ground reaction forces} (GRF) measurements gaining new knowledge, identify unseen pattern and recognize connections.
KAMAS~\cite{wagner_2017_kamas} is a similar knowledge-assisted VA environment for analyzing event sequences and categorizing suspicious sequences according to a taxonomy of behaviors. Even though KAMAS was originally designed for IT-security analysts, comparable data structures and analysis problems are also relevant for healthcare (e.g.,~\cite{wang_2011_extracting, wongsuphasawat_2012_exploring}).

\textbf{Summarizing these findings}, it can be seen, that most of the discussed approaches cover how explicit domain knowledge can be exploited to enhance visual representation and data analysis; some approaches provide methods to generate explicit knowledge.
Additionally, most of the currently implemented knowledge-assisted VA environments are focused on the integration of specific domain knowledge, which could only be used for precisely defined analysis tasks.
In general, explicit knowledge is now a first-class artifact in the VA process but its form and structure are left unspecified.
None of the presented approaches provides a structural framework for describing and storing explicit knowledge in VA environments.
Thus, a structural framework is needed and combined with the theoretical process model by Federico, Wagner et al.~\cite{federico-wagner_2017_model} it would provide valuable generative guidelines for the development of novel knowledge-assisted VA environments.

\section{Desiderata}
\label{sec:desiderata}

The following desiderata for a structural model of explicit domain knowledge were established in the reflective discussions of two recent design study projects,
in which our collaboration with domain experts resulted in a VA environment \cite{aigner_2018_kava,wagner_2017_kamas,wagner_2018_kavagait}
and preceding design study work that only sketched the integration of explicit knowledge
\cite{rind_2011_visualexploration, aigner_2012_comparative, federico_2014_qualizon}.
Additional inputs result from analysis of the scientific work cited above.
In particular, we build upon the characterization of knowledge by Federico, Wagner et al.~\cite{federico-wagner_2017_model} with its three axes space, type, and origin. Their knowledge characterization, however, is only descriptive about what is possible and is used to categorize knowledge-assisted VA environments for their survey.

Overall, we envision explicit domain knowledge as a first-class artifact in VA process that is both an input and an output of VA activities \cite{federico-wagner_2017_model, lammarsch_2011_towards}.
As analytics in the ``wild'' are seldom constrained to a single isolated VA environment or a single data backend \cite{fekete_2013_visual},
we regard it imperative to design a framework for explicit domain knowledge in a way to allow manifestation of knowledge in various data structures and backends as well as utilization in different VA environments. 

\newcommand{\desit}[1]{\textbf{#1}}
\newcommand{\deshe}[1]{}

The nine desiderata are:

\begin{enumerate}[D1:]
 \item \label{req:interpret}
\deshe{Machine Interpretability:}
Explicit knowledge, by definition, resides in the machine space
and is \desit{machine interpretable} \cite{federico-wagner_2017_model,wang_defining_2009}.
Structured forms of knowledge representation allow the VA environment to reason about relationships within the knowledge
(e.g., controlled vocabularies with same\slash{}different, taxonomies with hierarchical,
or ontologies with custom relationships).
While free text annotations or hand drawn polygons can capture knowledge in the machine space,
they are, in their raw form, opaque to machine reasoning.

 \item \label{req:domain}
 The structural model should focus on \desit{domain knowledge}, i.e. interpreting the data.
In contrast, operational knowledge, i.e.\ effectively using the VA environment,
is out of its scope.
While the latter can be tackled by usable user interfaces, user onboarding,
and automated visualization recommendation,
the former is essential for the success analysis, either as explicit knowledge
or as tacit knowledge of the domain experts as user \cite{federico-wagner_2017_model}.

 \item \label{req:preexisting}
It should be possible to reuse \desit{pre-existing taxonomies} or other knowledge artifacts
that have been established within a community of practice.

 \item \label{req:exchange}
The structural framework should facilitate the \desit{exchange} of explicit knowledge between different VA environments.

 \item \label{req:structure}
The structural framework should be compatible with \desit{heterogeneous data structures} and storage technologies.

 \item \label{req:origins}
\desit{Runtime editing} of explicit knowledge should be possible from within the VA environment.
Thus, all the origins of knowledge sketched by Federico, Wagner et al.~\cite{federico-wagner_2017_model} should be supported:
\begin{inparaenum}[\itshape (i)]
 \item knowledge artifacts can predate the VA environment (cp.\ D\ref{req:preexisting});
 \item explicit knowledge can be prepared during the VA environment's design process;
 \item a single user can interactively externalize domain knowledge;
 \item multiple users can externalize and share knowledge; or
 \item knowledge can be automatically derived from data \cite{federico-wagner_2017_model}.
\end{inparaenum}

 \item \label{req:provenance}
In order to support \desit{provenance} and accountability of knowledge generated, especially in the post-design phase,
the framework should provide a standardized form to include authorship details and further provenance information.

 \item \label{req:library}
There should be good \desit{software library support} to integrate explicit knowledge into the source code of a VA environment.
Widely used development languages like JavaScript, Python, and Java should be supported.

 \item \label{req:readable}
The notation of explicit knowledge should be easily \desit{readable} and editable by knowledge engineers and visualization designers for development and debugging.

\end{enumerate}

The last two desiderata (D\ref{req:library} and D\ref{req:readable}) might appear to be overly specific.
Of course, it is possible to rely on custom software solutions and knowledge editors.
However, our practical experience in designing knowledge-assisted VA environments underscored
the importance of suitable software development support and easily debuggable notation
to allow for the rapid feedback cycles typical in visualization design studies\cite{mccurdy_2016_action}.

\section{Structural Framework} 
\label{sec:model}

\begin{figure}[h]
\centering
\includegraphics[width=\columnwidth]{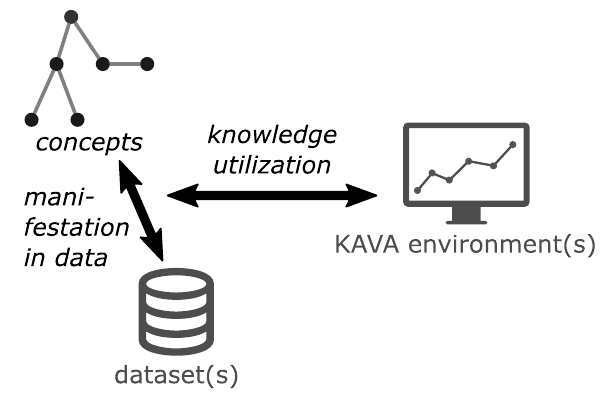}
\caption{The structural framework of explicit domain knowledge consists of concepts, their manifestation in datasets, and the utilization in KAVA environments.}
\label{fig:parts}
\end{figure}

The desiderata collected above characterize a generalized structural framework
that enables the communication and reuse of explicit domain knowledge
across different KAVA environments (D\ref{req:exchange}),
different datasets (D\ref{req:structure}), and different users (D\ref{req:origins}).
In order to separate these aspects clearly,
we suggest a structural framework consisting of three components (Fig.~\ref{fig:parts}).

\begin{enumerate}
 \item
Concepts from the application domain and their relationships.

\textit{Example (diabetes): the diagnosis `hyperglycemia', i.e.\ high blood sugar.}

\textit{Example (physiotherapy): a gait abnormality of the knee during mid-stance phase.}

 \item
A mapping between concepts and the dataset(s) to manifest knowledge in the relevant data items.

\textit{Example: a blood sugar level higher than 200 mg/dl is diagnosed as hyperglycemia.}

 \item
A mapping between concepts and the KAVA environment(s) to utilize the knowledge.

\textit{Example (diabetes): visually represent period of hyperglycemia as a horizontal line.}

\textit{Example (physiotherapy): provide list of known gait patterns and highlight patient marks on selection.}
\end{enumerate}

The vocabulary of concepts will remain comparatively stable for most domain problems (Table~\ref{tab:effects}).
Therefore, knowledge about concepts can act as an anchor when linking additional datasets to the analysis.
If these datasets are heterogeneously structured (D\ref{req:structure}), adaptations to the manifestation component will be needed.
Likewise, if additional VA environments are used (D\ref{req:exchange}), the knowledge utilization mapping may be adapted.
All three knowledge components can be manipulated by analysts via interaction with the VA environment (D\ref{req:origins}),
whereby we assume that most changes will affect the manifestation component.

\begin{table}
\begin{center}
\caption{Components of the structural knowledge framework and their dependence on domain problem, datasets, KAVA environments, and user interaction (\textbullet \dots high, $\circ$\dots low).}%
\label{tab:effects}
\begin{tabular}{r@{\quad}c@{\quad}c@{\quad}c}
\toprule
                    & concepts & manifestation & utilization \\
\midrule
domain problem      & \textbullet & $\circ$ & $\circ$ \\
dataset(s)          &   & \textbullet & \\
KAVA environment(s) &   &   & \textbullet \\
user interaction    & $\circ$ & \textbullet & $\circ$ \\
\bottomrule
\end{tabular}
\end{center}
\end{table}

\subsection{Knowledge about Concepts}

Concepts are in the center of the structural framework because they are comparatively stable over time.
Since the concept component of the structural framework is independent from datasets and KAVA environments,
it is possible to apply existing work from the field of knowledge representation.
Linked data~\cite{berners_2006_ld} and in particular the Resource Description Framework (RDF)~\cite{cyganiak_2014_rdf} are established approaches
for representing semantic information 
in a machine-interpretable form (D\ref{req:interpret}).

The RDF models information about concepts in a directed graph
that is specified as  a set of triples.
Each triple consists of a subject node, a property predicate, and an object, which can be a node or a literal.
A node is identified by a globally unique resource identifier and can refer to anything from concrete persons/things to an abstract concept.
For storing and exchanging these triples, several serialization formats exist.
In the context of KAVA, we adopt in particular the Turtle and JSON-LD formats.
Turtle \cite{beckett_2011_turtle} has a concise, text-based syntax
that is suitable for debugging by VA designers and knowledge engineers (D\ref{req:readable}).
JSON-LD \cite{sporny_2014_json-ld} is a JSON-based format
and, thus, compatible with many modern VA software libraries (D\ref{req:library}).
Alternatively, particular software libraries such as rdflib.js \cite{rdflib} directly support RDF and allow for operations based graph relations and semantics (D\ref{req:interpret}).

Often there will be a pre-existing concept schemes such as ICD-10, MeSH, or SNOMED CT \cite{dalianis_2018_medical}
that can be built upon (D\ref{req:preexisting}).\\
\textit{Example (diabetes): in the ICD-10 classification of health issues \cite{who_2016_icd}, there is a concept for hyperglycemia (without diabetes or other known diagnosis) in the R73 subbranch. In WikiData, hyperglycemia is known as concept Q271993 \cite{wiki_hypergly}.}

\begin{figure}[h]
\centering
\includegraphics[scale=0.25]{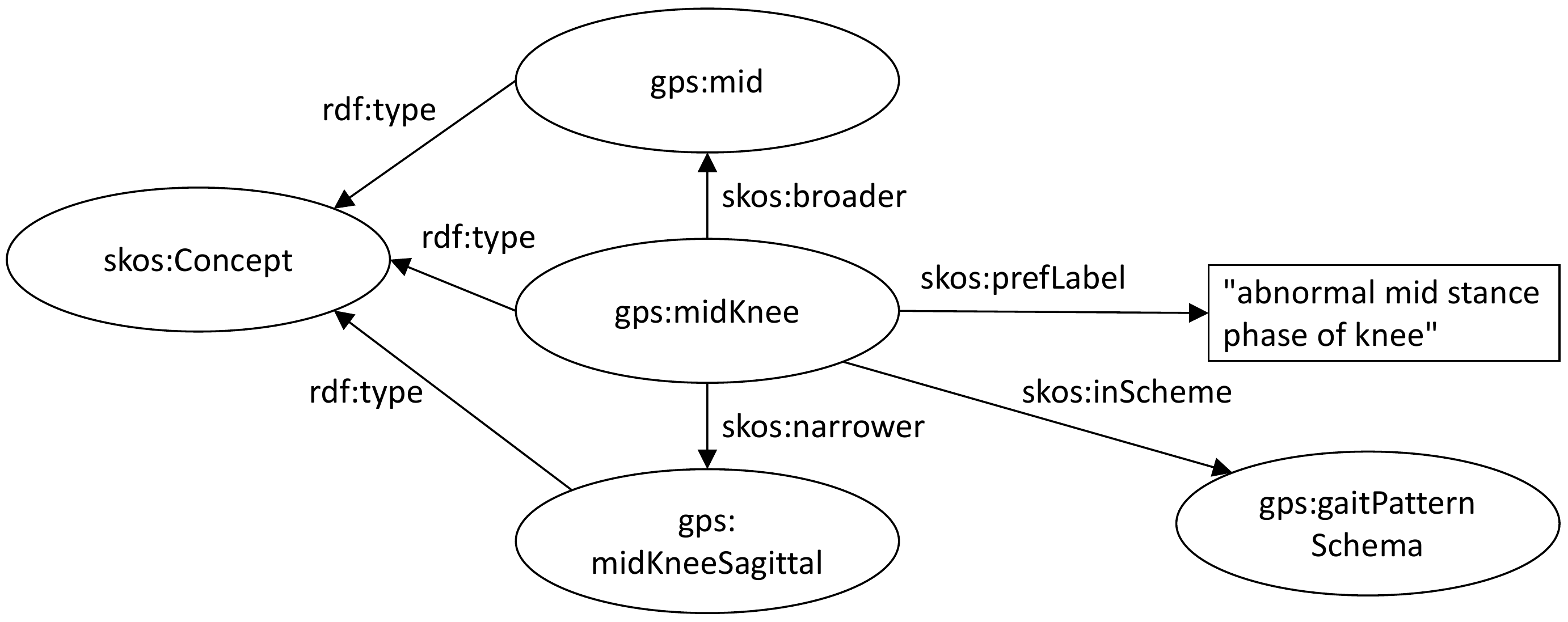}
\caption{A gait pattern concept described with SKOS in visual form.}
\label{fig:gait-concepts}
\end{figure}

If a custom concept scheme is needed,
the Simple Knowledge Organization System (SKOS) \cite{isaac_2009_skos} provides a compact RDF vocabulary
to describe the concepts of a semi-formal knowledge organization system such as a thesaurus or a taxonomy.
The vocabulary of SKOS includes labels to describe concepts in natural language, a broader\slash{}narrower relationship for hierarchical links between concepts,
and a related relationship for associative links.\\
\textit{Example (physiotherapy): a gait analysis laboratory characterizes abnormal gait pattern by concepts along a 3-level taxonomy that distinguishes first by gait phase (e.g., `mid stance'), then the affected joint (e.g., `knee'), and finally the direction (e.g., `sagittal') (Fig.~\ref{fig:gait-concepts}, Listings \ref{lst:skosttl} and \ref{lst:skosjson})}.

\begin{figure}[h]
\renewcommand{\thelstlisting}{\arabic{lstlisting}}
\begin{lstlisting}[caption=A gait pattern concept described with SKOS in Turtle format.\label{lst:skosttl}]
gps:midKnee rdf:type skos:Concept;
	skos:prefLabel "abnormal mid stance phase of knee";
	skos:broader gps:mid;
	skos:narrower gps:midKneeSagittal;
	skos:inScheme gps:gaitPatternSchema.
\end{lstlisting}

\renewcommand{\thelstlisting}{\arabic{lstlisting}}
\begin{lstlisting}[caption=The same gait pattern concept described with SKOS in JSON-LD format.\label{lst:skosjson}]
{
	"@id": "gps:midKnee",
	"@type": "skos:Concept",
	"skos:broader": {
		"@id": "gps:midKnee"
	},
	"skos:narrower": {
		"@id": "gps:midKneeSagittal"
	},
	"skos:inScheme": {
		"@id": "gps:gaitPatternSchema"
	},
	"skos:prefLabel": "abnormal mid stance phase of knee"
}, ...
\end{lstlisting}
\end{figure}

\subsection{Knowledge Manifestation in Datasets}
In addition to the concept component described above,
a structural framework of explicit  knowledge for VA needs a component
to manifest domain concepts in the respective data items,
because VA is primarily concerned with analysis of datasets (D\ref{req:domain}).
We model also this component of the knowledge framework as an RDF graph,
which makes it possible to leverage the semantics of concept
and two established approaches for provenance (D\ref{req:provenance}),
the Dublin Core and the FOAF vocabularies \cite{cyganiak_2014_rdf,isaac_2009_skos}.
The RDF properties necessary for manifestation of domain knowledge will be illustrated in this subsection.

\begin{figure}[h]
\begin{center}
\includegraphics[scale=0.25]{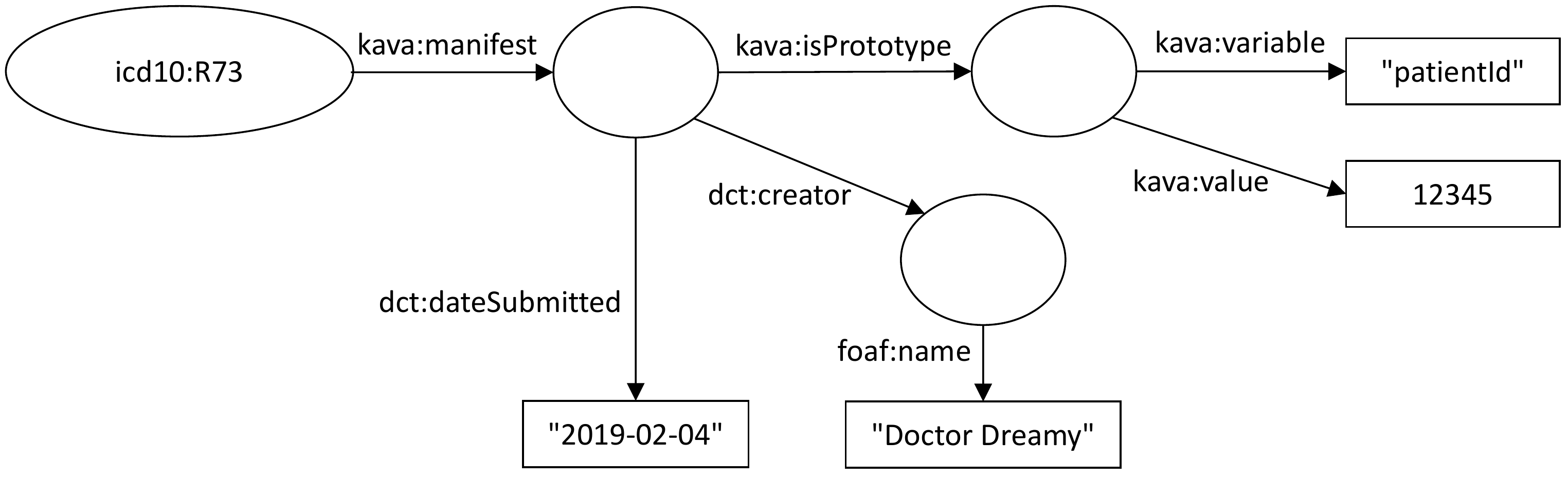}
\caption{A health concept manifested on data by a prototypical patient.}
\label{fig:health-embed-proto}
\end{center}

\begin{lstlisting}[caption=The hyperglycemia concept manifested on a patient's data by direct mapping.\label{lst:health-embed-proto}]
icd10:R73 kava:manifest [
	kava:isPrototype [
		kava:variable "patientId";
		kava:value 12345
	];
	dct:creator [foaf:name "Doctor Dreamy"];
	dct:dateSubmitted "2019-02-04"
].
\end{lstlisting}

\end{figure}

The concepts can be manifested on dataset items through either a direct or an indirect mapping
 depending on whether references or characteristics are given \cite{andrienko_2006_exploratory}.
A \emph{direct mapping} annotates individual known occurrences of a concept in the datasets by specifying their references, i.e. values for their identifying variables.
Such explicit knowledge can be utilized
\begin{inparaenum}[\itshape (i)]
 \item
to train classification models from the annotated prototypes,
 \item
to reconfirm the results of such models,
 \item
to interpret data in context \textit{(Example in healthcare: blood sugar levels during pregnancy.)},
or even
 \item
to hide irrelevant parts of the data.
\end{inparaenum}
\\
\textit{Example (diabetes): the patient with ID 12345 is marked as having hyperglycemia.
This knowledge has provenance to be created by `Doctor Dreamy' on Feb 4, 2019 (Fig.~\ref{fig:health-embed-proto} and Listing~\ref{lst:health-embed-proto}).}

\begin{figure}[h]
\begin{lstlisting}[caption=The hyperglycemia concept manifested by indirect mapping with a threshold on a blood test.\label{lst:embed-health-indirect}]
icd10:R73 kava:manifest [
	kava:matchVariable [
		kava:variable health:bloodSugar;
		kava:minValue 200
	]
].
\end{lstlisting}

\begin{lstlisting}[caption=The hyperglycemia concept manifested by indirect mapping with a custom query string on blood laboratory data.\label{lst:embed-health-custom}]
icd10:R73 kava:manifest [
	kava:matchQuery "[glucose] > 200";
	dct:creator [foaf:name "ACME laboratory equipment"]
].
\end{lstlisting}
\end{figure}

An \emph{indirect mapping} describes the characteristics of all occurrences of a concept rather than identifying individual exemplars.
The characteristics can be specified in a query predicate.
Typical query predicates, e.g., on multivariate tables or sequences, can be expressed in a structured way by an RDF vocabulary.
Yet, to support the widest possible range of dataset structures and data storage technologies (D\ref{req:structure}), the query predicate can also be given as a string using the query language of the underlying technology (e.g., SQL, XQuery, SPARQL, Gremlin).
Reusing query strings of existing languages will also reduce the effort for developers (D\ref{req:readable}).
\\
\textit{Example (diabetes): hyperglycemia can be characterized by blood sugar test over 200mg/dl (Listings~\ref{lst:embed-health-indirect} and \ref{lst:embed-health-custom}).}

\subsection{Knowledge Utilization} 

Finally, concepts and their manifestation in data need to be utilized in the VA environment.
A solid starting point for the utilization component are existing visualization grammars
such as Vega-Lite~\cite{satyanarayan_2017_vega-lite},
which describes data transformations, encoding on visual marks, and interactivity in a declarative JSON-based format.

However, there is a wide range of possible knowledge utilizations
and these are integrated deeply into the application logic of their VA environments.
Therefore, this component of the structural knowledge framework might be least amendable for generalization.
Some typically approaches will be illustrated below.

\begin{itemize}
 \item
All relevant concepts can be visualized as marks.
It is possible to create a tree visualization based on broader/narrower relationships in SKOS 
or a network visualization based on the related relationships.
Visual encoding channels can indicate how frequent a concept is in the data 
or how similar prototypical items of the concept are to the currently analyzed data (e.g., KAVAGait Fig.~\ref{fig:kavaGaitInterface}.1.a).
 \item
The marks of data items can have a different visual encoding depending on their manifested knowledge concepts -- mapped either directly as a prototype or indirectly by fulfilling a query predicate. 
 \item
Multiple data items manifested with the same concept can be shown as an aggregate mark (e.g., a horizontal line spanning the time period, while a patient suffered from hyperglycemia).
 \item
Query predicates can be parsed and visualized.
For example, the area above 200mg/dl could be colored in the background of a blood sugar line plot to indicate risk of hyperglycemia.
\end{itemize}

\section{Applying the Framework in KAVAGait}
\label{sec:case}

Next, we demonstrate how the explicit knowledge of an exemplary VA environment can be structured according to our framework.
This scenario is based on the KAVAGait \cite{wagner_2018_kavagait} design study with physiotherapists
and illustrated using the processes of the knowledge-assisted VA model \cite{federico-wagner_2017_model,wagner_2017_integrating}.

\begin{figure*}[ht!]
  	\centering
  		\includegraphics[width=1\textwidth]{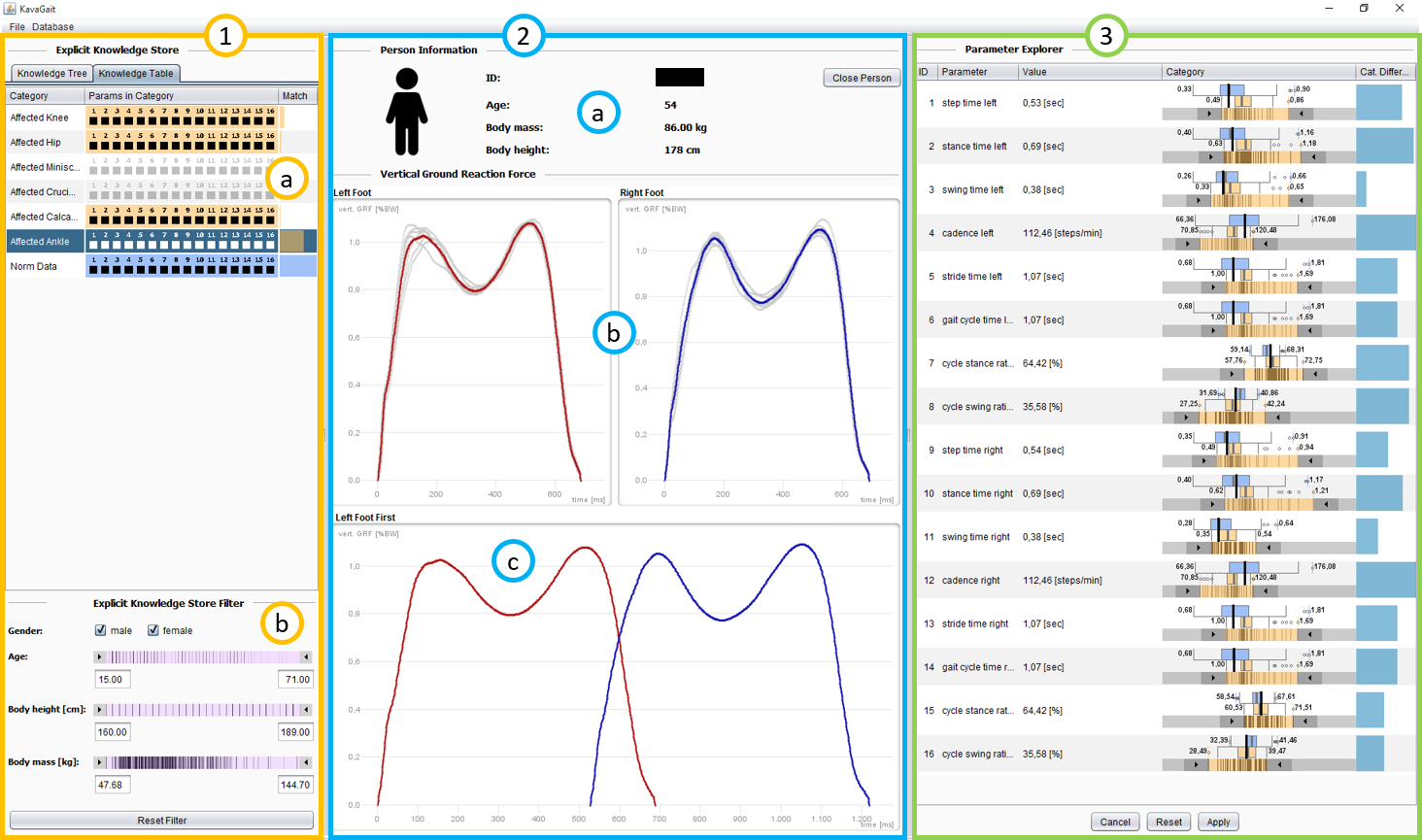}
  	\caption[Interface of KAVAGait]{\textbf{Interface of KAVAGait} -- User interface of the KAVAGait prototype with its three main areas for gait analysis~\cite{wagner_2018_kavagait}.
    (1) The \highlight{Explicit Knowledge Store} shows (1.a) a table with the patient's match for different gait categories and (1.b) allows filtering of the population of prototypical patients.
    (2) The patient explorer including the (2.a) \highlight{Person Information}, the (2.b) visualization of the ground reaction force (${F_{v}}$)  time series for each foot on a separated scale and the (2.c) visualization of the combined ${F_{v}}$ from both feet.
    (3) Shows the \highlight{Parameter Explorer} visualizing the 16 calculated \highlight{Spatio-Temporal Parameters} of the loaded patient in relation to the \highlight{Norm Data Category} and a second \highlight{Selected Category}.
    \newline \textit{Image courtesy of~M.\,Wagner\,\cite{wagner_2017_integrating}.}
    }
	\label{fig:kavaGaitInterface}
\end{figure*}

\begin{figure}[h!]
\centering
  \includegraphics[width=\columnwidth]{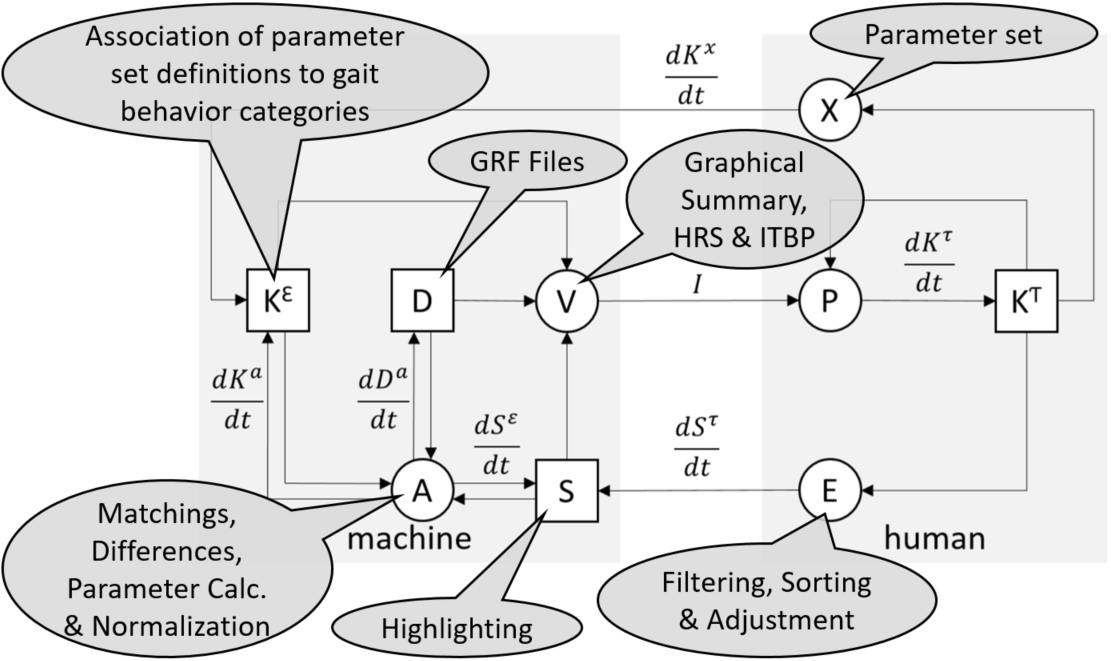}
  \caption[Instantiation of Knowledge-assisted VA Model for KAVAGait]{\textbf{Instantiation of Knowledge-assisted VA Model for KAVAGait} -- Illustrating the different prototype specific elements in relation to the related components and processes of the\ \highlight{Knowledge-assisted VA Model}~\cite{federico-wagner_2017_model,wagner_2017_integrating} (Important abbreviations included in the green bubbles: GRF := ground reaction force, HRS := hatching range-slicer, ITBP := interactive twin-box-plot).
  \hfill \textit{Image courtesy of~M.\,Wagner\,\cite{wagner_2017_integrating}.}
  }
    \label{fig:kavaGaitModel}
\end{figure}

\textbf{KAVAGait}~\cite{wagner_2018_kavagait} is a \highlight{Knowledge-assisted VA System for Clinical Gait Analysis}, whereby the analysts (clinicians) are supported during analysis and clinical decision making (see Fig.~\ref{fig:kavaGaitInterface}). 
The analysts have the ability to load patient gait data containing time series of ground reaction force measurements for each foot.
The time series are visualized as line plots in the center of the user interface, describing the force over time. Additionally, 16 spatio-temporal parameters (e.g., step time, stance time, cadence) related to the loaded patient's gait are calculated, visualized, and used for automated patient comparison and categorization based on the introduced interactive twin box plots.
One primary goal during clinical gait analysis is to assess whether a recorded gait measurement displays \highlight{normal gait} behavior or if not, which specific \highlight{gait abnormality} is present. Thus, the environment's explicit knowledge store contains several categories of \highlight{gait abnormalities} (relating to e.g., knee, hip, ankle) as well as a category including \highlight{healthy gait pattern} data, used for analysis and comparison by default.
Each category is described by the data of patients that were previously assigned to this category.
In particular, the $[min, max]$ ranges of the patients' 16 spatio-temporal parameters are calculated.
Based on these category descriptions, automated data analysis of newly loaded patient data is provided (e.g., automatically calculated category matching). This automated data analysis supports the analysts in their interactive data exploration.
To achieve a second goal, clinicians can generate new explicit knowledge by adding the analysis result to the explicit knowledge store.
KAVAGait also provides the ability to interactively explore and adjust the internally stored explicit knowledge.

\textbf{Assessment of Patient Data:} If the analyst loads a gait analysis file $D$ of a patient into KAVAGait (see Fig.~\ref{fig:kavaGaitModel} for a graphical overview of the knowledge processes), first, the contained time series are visualized $V$ based on the systems specification $S$.
Second, the stored explicit knowledge $K^\epsilon$ and the automated data analysis methods $A$ are strongly intertwined with all components of the VA environment. Thus, this pipeline immediately calculates the 16 spatio-temporal parameters based on the loaded time series and the matching to the different knowledge categories, affecting the specification.
The combination of both former described procedures can be expressed as the initial analysis and visualization pipeline. However, if the specification is not influenced by the analyst (e.g., zooming, filtering, sorting), all stored explicit knowledge is used for analysis and comparison.
Based on the generated visualization, the image $I$ is perceived by the analyst, gaining tacit knowledge $K^\tau$, influencing the analysts perception $P$. 
Depending on the tacit knowledge, the analyst has now the ability to interactively explore $E$ the visualized time series and spatio-temporal parameters by using the environment's provided methods (e.g., zooming, filtering, sorting).
During this iterative process, the analyst gains further tacit knowledge based on the adjusted visualization.

\textbf{Explicit Knowledge Generation and Adjustment:}
To generate explicit knowledge $K^\epsilon$ (see Fig.~\ref{fig:kavaGaitModel} for a graphical overview), the analyst has the ability to include the spatio-temporal parameters of analyzed patients based on his/her clinical decisions to the explicit knowledge store, which can be described as the extraction $X$ of tacit knowledge. This explicit knowledge can be visualized in a separated view, whereby the explicit knowledge is automatically transformed into a dataset. Different views are providing the adjustment of the stored explicit knowledge by the analyst's tacit knowledge.

\textbf{Structuring the Explicit Knowledge:}
KAVAGait operates with explicit domain knowledge about different gait patterns.
Thus, it needs a concept for each pattern.
While a multi-level taxonomy of gait patterns would be possible (cp.\ Listing~\ref{lst:skosttl}),
for the KAVAGait prototype a flat list of 7 concepts is sufficient (`affected knee', \dots, `norm data').
The explicit knowledge is manifested in $[min, max]$ ranges for the spatio-temporal parameters.
Therefore, KAVAGait needs a two-part approach:
(1) Typically, the clinician categorizes the gait pattern of a patients and includes them in the explicit knowledge store under a category.
In the explicit knowledge framework, this is expressed as a direct mapping linking the concept to their prototypical patients (cp.\ Listing~\ref{lst:health-embed-proto}).
The $[min, max]$ ranges need to be calculated dynamically from the prototypical patients' spatio-temporal parameters,
because the patient population can be filtered e.g., by age.
(2) If the clinician manually adjusts a $[min, max]$ range, this can be expressed as an indirect mapping with the given value range  (cp.\ Listing~\ref{lst:embed-health-indirect}).
This explicit knowledge is utilized in multiple ways:
(i) the concepts are represented as rows in the \highlight{Knowledge Table} (Fig.~\ref{fig:kavaGaitInterface}.1.a).
The values of the prototypical patients' spatio-temporal parameters are shown as hatches in the \highlight{Parameter Explorer} (Fig.~\ref{fig:kavaGaitInterface}.3) and to calculate the matchings and category differences.
The $[min, max]$ ranges, which are either calculated from prototypical patients or manually adjusted, are visualized by the range sliders in the \highlight{Parameter Explorer} and the boxes of the \highlight{Knowledge Table}.

\textbf{Summary:} This design study demonstrates that explicit knowledge extracted from the clinicians tacit knowledge opens the possibility to support clinicians during clinical decision making. Additionally, KAVAGait could also be used to share the knowledge of domain experts as well as to use it for educational support. 
In contrast to other analysis systems (e.g., based on MatLab), KAVAGait uses analytical and visual representation methods to provide a scalable and problem-tailored visualization solution following the visual analytics agenda~\cite{thomas_2005_illuminating,keim_2010_mastering}.
For keeping up with the large number of patients stored as explicit knowledge, clinical gait analysts need to continuously adapt the systems settings during the clinical decision making process. Supporting such interactive workflows is a key strength of visualization systems. Clinical gait analysis in particular profits from extensive interaction and annotation because it is a very knowledge-intensive job.
By providing knowledge-oriented interactions, externalized knowledge can subsequently be used in the analysis process to support the clinicians. 
The newly developed visual metaphors provide an easy way to inspect variability of the data (e.g., standard deviation), allow to identify outliers in the data, and provide an easy to understand overview of the data and automated matching results. Additionally, based on the interactive twin box plots, it is possible to perform intercategory and patient comparisons by details on demand to find similarities in the data.

\section{Conclusions and Next Steps}

Addressing the need for deeper integration of domain knowledge,
existing VA environments have found a multitude of mechanisms to structure and manage explicit domain knowledge.
This paper set out to reflect and generalize the results from existing design study projects
and work towards a structural framework for explicit domain knowledge in VA.
This reflection has identified nine desiderata for such a structural framework.
A preliminary structural framework is proposed that separates the concerns of explicit knowledge into three components: concepts, manifestation, and utilization.
For the components concepts and manifestation, we apply linked data using SKOS and a custom RDF vocabulary for direct and indirect mapping to data items.
The utilization component can be addressed based on the JSON-based Vega-Lite language.
Being limited to the retrospective reflection of several knowledge-assisted VA design studies,
this preliminary structural framework lacks any claims of completeness.
Notwithstanding its limitations, this work certainly adds to our understanding of explicit knowledge in VA environments and provides a frame of reference for future design studies.

A natural progression of this work is to analyze a larger sample of VA environments in more detail by reverse engineering their explicit knowledge using this theoretical framework.
Such a structured literature review can both survey the landscape of knowledge-assisted VA environments and identify missing vocabulary for knowledge manifestation in heterogeneous datasets.
In particular, further modelling needs to examine more closely
how concrete variables in datasets can be group to abstract concepts (e.g., a concept for blood sugar).
In addition, the scope of knowledge utilization approaches can be more comprehensively surveyed.
As a next step, the emerging structural knowledge framework needs to be tested in practice,
either in new design study project or through evolution of an existing VA environment.
Given the general claim, this implementation can also result in a reusable software library.
Finally, the aspired real-world utility can only be assessed through empirical studies with users of VA environments.

\acknowledgments{
This work was partly funded
by the Austrian Science Fund (FWF): P25489-N23 via KAVA-Time
and by the Austrian Research Promotion Agency (FFG): grant \#866855.
}

\bibliographystyle{abbrv-doi-narrow}

\bibliography{ref-knowledge}
\end{document}